\documentstyle[aaspp4]{article}

\lefthead{Feintein, Ib\'a\~nez \& Ruth Lazkoz}
\righthead{Disks Expanding FRW Universes} 
\begin{document}

\title{Disks in Expanding FRW Universes}

\author{A. Feinstein, J. Ib\'a\~nez and Ruth Lazkoz}
\affil{Dpto. F\'{\i}sica Te\'orica, 
Universidad del Pa\'{\i}s Vasco, \\
Apdo. 644,
E-48080 Bilbao, Spain; \\wtpfexxa@lg.ehu.es,
 wtpibmej@lg.ehu.es and wtblasar@lg.ehu.es}

\begin{abstract}
We construct exact solutions to Einstein equations which represent
relativistic disks imbedded in an expanding FRW Universe. It is 
shown that the expansion influences the kinematical characteristics 
of the disks such as rotational curves, surface mass density, etc.
 The effects of the expansion are exemplified with non-static 
generalizations of Kuzmin-Curzon and Schwarzschild disks.
\end{abstract}


%-------------------------------------------------------------------%
%-------------------------------------------------------------------%
%---Axisymmetric solutions of the Einstein field equations----------%
%-------------------------------------------------------------------%
%-------------------------------------------------------------------%
\section{Introduction}
Recently, there has  been a renewed interest in the study of
 relativistic disks (Bi\u{c}\'ak, Lynden-Bell \& Katz, 1993;
Bi\u{c}\'ak  \& Ledvinka, 1993; \& Bi\u{c}\'ak, Lynden-Bell 
\& Pichon, 1993), stemming from a desire to address
 the possible observational  evidence of giant black holes
 sustained by the surrounding disks. The gravitational fields of 
galactic disks may be accurately modeled by the Newtonian potential
 theory  provided the thickness of the disk is negligible compared 
with the typical size of their halos. In more sound situations, 
however, when gravity is strong enough one must turn to General
Relativity. A typical situation, where it is believed that Einstein
theory may contribute is met when one considers the accretion
 disks around the central black holes in quasars. Disks may be also 
used to model sheet-like structures, so their study may shed some 
light on the understanding of the large scale inhomogeneities present 
in the Universe (\cite{lemos94}).

In General Relativity, solutions representing thin disks may be 
constructed starting with static axially symmetric vacuum Weyl
metrics. A discontinuity is then introduced in the first 
derivative of the metric across the azimuthal plane $z=0$, which
in turn  induces distribution-like terms in the Ricci tensor. 
Everywhere outside the $z=0$ plane the solution is vacuum with a 
non-vanishing surface mass density concentrated on this plane.
Since Weyl solutions are static, they should represent
static disks. To allow for rotation in General Relativity one must 
drop the assumption of the orthogonality of the two Killing vector 
fields associated with Weyl geometry, thus complicating the problem by
introducing the dragging of inertial frames.

Fortunately, Morgan and Morgan (\cite{morgan69a}; \cite{morgan69b};
 \cite{morgan70}) had the very smart
idea of interpreting the ``static disks" as made out  
of two equal streams of collisionless  particles circulating in
opposite directions. Consequently, the total angular momentum vanishes and the
system may be described by a Weyl line element. 
Interestingly enough, there exists observational evidence of some 
galaxies with two counterrotating stellar components around their 
center (\cite{rix92}; \cite{konrad94}; \cite{bertola96};
 \cite{konrad96}).

Another interpretational problem with the solutions described
by the Weyl metric arises due to the fact that the disks are 
infinite in their extent.  One may argue, however, that infinite
disks  model the inner portions of galaxies or accretion disks.
Once the interpretational problems are dealt with, it 
is not difficult to model relativistic disks and 
study their physical properties such as velocity profiles, surface
 mass density, redshifts and so on.

Evans and de Zeeuw (\cite{evans92}) showed that it is possible
to analyze any classical axially symmetric disk into a
linear superposition of so-called Kuzmin disks (\cite{kuzmin56}).
Using this result, Bi\u{c}\'ak, Lynden-Bell and Katz (\cite{bijak93},
 hereafter BLK) constructed families of counterrotating disk 
space-times. Their work was further extended by Bi\u{c}\'ak, 
Lynden-Bell and Pichon (\cite{pichon93}, hereafter BLP) and more
 recently by Lynden-Bell and Pichon (\cite{pichon96}).
On the other hand, Lemos and Letelier (\cite{letelier93};
 \cite{letelier94}; \cite{letelier96}) using the well known techniques
 of superposing different Weyl solutions, have constructed
 space-times describing disks surrounding  
Schwarzschild black holes. 

The main purpose of this paper is to consider, in the 
framework of exact solutions in General Relativity, the
 effect of the cosmological expansion on the disk dynamics on one 
hand, and the influence of disk-like large-scale structures on the 
cosmological model's kinematics on the other. It is known
 (\cite{petros71}) that the effects of the expansion on bounded 
gravitational systems is proportional to the ratio $\bar \rho/\rho_s$,
 where $\bar \rho$ is the background energy density and $\rho_s$ is the
 mean rest-density of the system. Therefore, for galactic
 and accretion disks fueling the quasars these effects are not
 expected to be too significant, though the overall expansion may
 change the disk dynamics. For disk-like 
large-scale structures, however, the expansion may turn out to be of
 particular importance.  

In \S 2 we propose an algorithm to generate solutions
to Einstein equations which may be interpreted as inhomogeneities
imbedded in a spatially flat isotropic Universe. The technique
 uses a scalar field generalization in a cosmological setting of
 static Weyl solutions. The scalar field can be split into two 
parts, one homogenous, which may be hydrodynamically interpreted
 as the velocity potential for an adiabatic perfect fluid acting as the
 source of a FRW expansion; and another one highly irregular which
 may be thought of as a local disk inhomogeneity. In \S 3 we apply the 
generating procedure of the previous section to 
the general relativistic  Kuzmin disks. We then analize the 
energy-momentum  tensor for the obtained solutions and interpret 
them as disks  imbedded  in a FRW background. \S 4 is devoted to
 study the dynamics of the disks with the emphasis on the effects 
of the expansion on pertinent dynamical and kinematical quantities.
We conclude the paper with the discussion and future prospects.

\section{Exact non-static solutions with axial symmetry}

\subsection{The generating algorithm}
In this Section we obtain non-static solutions to Einstein field
equations representing compact objects in a cosmological
setting. We start with Weyl's line element which  can be written 
as:
 \begin{equation} 
ds^2 =-e^{2\,\nu(\rho ,z)} {dt}^2+e^{-2\,\nu(\rho ,z)} \rho^2
d{\phi}^2+ e^{-2\,\nu(\rho,z)+2\,\zeta \,(\rho,z)} (
d{\rho}^2+dz^2)\,.\label{metric}
\end{equation}
Throughout the paper we use $G=c=1$. It is well-known that vacuum
 static axially symmetric fields in 
General Relativity can be generated starting with a
 Newtonian potential (\cite{weyl17}; \cite{levi19a}; \cite{levi19b}). 
For the space-times obtained in this way the metric function
$\nu$ in equation $(\ref{metric})$ may be taken to be any classical
solution of the Laplace equation in cylindrical coordinates,
 and the other metric function $\zeta$ is obtained by a quadrature.

New non-static metrics may be  generated in two stages. First, 
starting from a vacuum solution of the Einstein equations for 
a metric given by equation $(\ref{metric})$, we construct a new 
static solution with a minimally coupled massless scalar field
 $\psi$ as a  source. For these space-times the energy-momentum 
tensor takes the form:
\begin{equation} 
T_{ab}={\psi}_{,a}{\psi}_{,b}-\frac{1}{2}\,g_{ab}
\,{\psi}_{,c}{\psi}^{\, ,c},
\end{equation}
and the set of Einstein equations is
 \begin{eqnarray} 
{\nu}_{zz}+{\nu}_{{\rho}{\rho}}+{\rho}^{-1}{\nu}_{{\rho}} =0\,,
 \label{ein1}\\
{\zeta}_{zz}+{\zeta}_{{\rho}{\rho}}-{\rho}^{-1}{\zeta}_{{\rho}}+ 
2\,{{\nu}_{\rho}}^2=- {{\psi}_{\rho}}^2\,, \label{ein2}\\
{\zeta}_{zz}+{\zeta}_{{\rho}{\rho}}+{\rho}^{-1}{\zeta}_{{\rho}}+ 
2\,{{\nu}_z}^2=- {{\psi}_z}^2\,, \label{ein3}\\
 {\rho}^{-1}{\zeta}_z-2\,{\nu}_{\rho}\,{\nu}_z=
-{\psi}_z {\psi}_{\rho}
\label{ein4}\,,
\end{eqnarray}
along with the Klein-Gordon equation
\begin{equation}
{\psi}_{zz}+{\psi}_{{\rho}{\rho}}+
{\rho}^{-1}{\psi}_{{\rho}} =0\label{ein5}\,.
\end{equation}
It is easy to see that if $\nu_o$ and $\zeta_o$ solve the vacuum 
Einstein equations, a solution to equations 
(\ref{ein1}-\ref{ein5}) is then given by
\begin{eqnarray}
\nu&=&\nu_o+C \log \rho\,, \\
\zeta&=&B\,{\zeta}_o+E\,\nu_o+F\,\log \rho\,,\\
\psi&=& A\,\nu_o+D \log \rho\,,
\label{newmet}
\end{eqnarray} 
where the
constants are subject to the following constraints:
 \begin{eqnarray}
2\,C+A\,D &=& E,  \\ A^2+2 &=& 2\,B\,,\\
 D^2+2\,C^2 &=& 2\,F\,. 
\end{eqnarray}
For the purposes of this work we will discard the logarithmic 
terms by setting the constants $C, F \hbox{ and }D$ to zero, 
ensuring thus asymptotic flatness at spatial infinity
in a static case. 

Once the solution with a massless scalar field metric has been
 constructed, we transform it into a non-static solution with a
 self-interacting scalar field with an exponential potential. This
 is accomplished by using an algorithm originally due to Fonarev
 (\cite{fon95}); see as well the generalization by Feinstein,
 Ib\'a\~nez \& Lazkoz (\cite{fein95}). The new solution in a 
synchronous system of coordinates reads
\begin{equation}
ds^2 =  -e^{2\,\nu(\rho ,z)} {dt}^2+R^2 (t)\,e^{-2\,\nu(\rho
,z)}\left({\rho}^2 d{\phi}^2+
e^{2\,\zeta \,(\rho,z)} ( d{\rho}^2+dz^2)\right )\,,
\label{generic}
\end{equation}
where the ``scale factor" is $R(t)=t^{2/k^2}$ and the new metric
functions are
\begin{equation}
\nu=\nu_o\quad\hbox{ and } \quad \zeta=(1+2/{k^2})
\,{\zeta}_o\,,
\end{equation}
the constant $k$ being  the slope of the potential 
$V=\Lambda\,e^{-k\,\psi}$ and
 $\Lambda={(12-2\,k^2)/k^4}$.
 The line element in equation $(\ref{generic})$ is a solution of the 
Einstein field equations with the energy momentum tensor given by
\begin{equation}
T_{a b}={\psi}_{,a}{\psi}_{,b}-g_{a b}
\left(\frac{1}{2}{\psi}_{,i}{\psi}^{,i}+\Lambda\,e^{-k\,\psi}\right)\,.
\end{equation} 

Furthermore, the new scalar field $\psi(t,\rho,z)$ splits into a
 homogeneous  and an inhomogeneous part, 
$\psi(t,\rho,z)=\psi_{h}+\psi_{inh}$ with $\psi_{h}$ and 
$\psi_{inh}$ given by
\begin{equation}
\psi_{h}=\frac{2}{k}\,\log t \quad\hbox{ and } \quad
\psi_{inh}=\frac{2}{k}\,\nu_o\,.
\end{equation}
Note that in the particular case $k^2=6$ the potential term vanishes,
 and one is left with a massless scalar field. 

\subsection{Interpretation of the new solutions}

As long as the seed static metric is asymptotically flat,
 the newly generated non-static solution is
asymptotically homogeneous and isotropic, and at spatial 
infinity the geometry is represented by the FRW metric 
\begin{equation}
ds^2 = - {dt}^2+t^{4/k^2}\left({\rho}^2 d{\phi}^2+ d{\rho}^2+
dz^2\right)\,. \label{frw}
\end{equation} 
We now identify the homogeneous part of the scalar field
  with the velocity potential of an irrotational perfect fluid
 which is the source of the  metric in equation (\ref{frw}). 
The energy-momentum  tensor is then
\begin{equation}
T_{ab}^{FRW}=(p+\mu)u_{a}u_{b}+p\,g_{ab}\,,
\end{equation}
where the four-velocity of the fluid is given by 
 \begin{equation}
u_{a}={{\psi}_{h,a}{\delta}_t^{a}\over\sqrt{(-{\psi}_{h,t} 
\,{\psi}^{h,t})}}\,,
\end{equation}
so that the pressure $p$ and the energy density $\mu$ become
\begin{equation}
\mu=-\frac{1}{2}\,{\psi}_{h,t}\,{\psi}^{h,t}+\Lambda\, t^{-2}
\quad\hbox{ and }\quad
p=-\frac{1}{2}\,{\psi}_{h,t}\,{\psi}^{h,t}-\Lambda \,t^{-2}.
\end{equation}
Substituting the expression for the scalar field we readily get
\begin{equation}
p=\frac{4\,k^2-12}{t^2\,k^4}\quad\hbox{  and  }\quad
\mu={12\over {t^2\,k^4}}\,.\label{baro}
\end{equation}
It is straightforward to see that equation  (\ref{baro})  
defines  the barotropic equation of state 
\begin{equation}
p=\gamma\,\mu  \quad \hbox{  and  }\quad \gamma=\frac{k^2-3}{3}\,.
\end{equation}

Note, however, that the scalar field has also an 
inhomogeneous component. Although the inhomogeneities
 steadily dilute as the spatial distance
 increases, they cannot be regarded to be negligible in the 
intermediate regions.

 It is well-known that as long as the gradient of the scalar
 field is timelike the energy momentum 
 tensor  can be interpreted in terms of a perfect fluid. The 
asymptotic homogeneity of the solution 
ensures that at any time the gradient of the scalar field 
is timelike outside a certain closed spatial region,
depending basically on the mass and compactness of the disk,
as well as on the strength of the scalar
field. Outside this region the perfect fluid interpretation
holds, and at spatial infinity
the fluid becomes homogeneous and isotropic as mentioned above. 

On the other hand, the evolution  of the local 
inhomogeneities with time may also be deduced 
 by studying the gradient of the scalar field:
 \begin{equation}
\psi_{,c}\psi^{,c}=\psi_{h,\,c}\psi^{h,\,c}+\psi_{inh,\,c}
\psi^{inh,\,c}\,,
\end{equation}where
\begin{equation}
\psi_{h,\,c}\psi^{h,\,c}=-\frac{4}{3\,\gamma+3}\,t^{-2}e^{-2\,\nu}
\label{grad1}\,,
\end{equation}
and
\begin{equation}
\psi_{inh,\,c}\psi^{inh,\,c}=\frac{4}{3\,\gamma+3}\,
t^{-\frac{4}{3\,\gamma+3}}e^{2\,\nu-2\,\zeta}
 (\nu_{,\rho}^2+\nu_{,z}^2)\,.
\label{grad2}
\end{equation}

 We have checked numerically that for the models with accelerated
expansion $\gamma <-1/3$ the outer region, where the homogeneous term 
of the scalar field dominates the one representing the spatial
 gradients, grows with time. On the other hand, the highly irregular 
inner region shrinks to zero. This behaviour is similar to that 
one found  in the inflationary scenario, in the sense that the
small-scale inhomogenities are washed away due to the accelerated
 expansion.

The highly inhomogenous region surrounding the
 disk plane where the perfect fluid interpretation
 is not valid may be thought of as playing the role of a cushion
 between the disk and the cosmological
 fluid, representing a transition region between both regimes.

We suggest, therefore, to interpret the solutions of the type given 
by equation (\ref{generic}) as describing an inhomogeneity embedded in 
an expanding Universe on the grounds that on large scales one recovers
 the Friedmannian behaviour while the local structure for small 
distances is governed by the seed metric. This interpretation 
will be re-enforced in the light of the results of following
 sections.
 
%--------------------------------------------------%
%--------------------------------------------------%
%--------Relativistic disks -----------------------%
%--------------------------------------------------%
%--------------------------------------------------% 
\section{Relativistic expanding thin disks}
We now apply the described algorithm to the
 well-known family of axially symmetric solutions representing
 infinitesimally thin disks in a relativistic context and
 construct  their dynamical counterparts. We are mostly interested
 in the influence of  the expansion on the kinematical quantities that
 characterize the disks. It is expected that for an expanding
 Universe the mass-energy density on the disk plane will 
steadily decrease as time goes by. However, the effect on other 
quantities such as velocity or angular momentum is not intuitively 
foreseeable.  In addition, we believe that the conclusions reached
 in this work can bring a new perspective onto the problem of
embedding irregular sources into standard cosmologies.

\subsection{Generation of thin disk families}
 The disk configurations  we are dealing with were firstly studied 
by Morgan \& Morgan (\cite{morgan69a}, \cite{morgan69b},
 \cite{morgan70}) and describe counterrotating  disks with the 
same number of leftwise and rightwise rotating particles. 
As was pointed out in  the previous section one can start with the
 Newtonian potential produced by a thin disk to obtain  its 
relativistic version. 

For the sake of simplicity, unless otherwise stated, we will 
focus the analysis on two families of disks,
 the Kuzmin-Curzon disks
 and the generalized Schwarzchild disks (BLK, 1993).
  The Kuzmin-Curzon disks may be used as building
blocks for infinite families of more complicated disk-like solutions;
 one should just superpose elementary Kuzmin-Curzon disks with 
different compacticities weighted by a function $W(b)$, as shown by
 Evans \& de Zeeuw (1992). Their work on Newtonian axisymmetric 
potential-density pairs was extended to a relativistic case by BLK, 
who gave the general form of the metric for a superposition of 
Kuzmin-Curzon disks.  

BLP gave a step further in the study of relativistic disks by
constructing the more involved families of relativistic versions of 
the Kuzmin-Toomre (\cite{toomre63}; \cite{nagai76}; Evans \& de 
Zeeuw 1992) and Kalnajs-Mestel disks (\cite{mestel63}; 
\cite{kalnajs76}). The latter family of solutions is particularly
interesting because they have long and flat rotation curves for 
certain values of the parameters, and include the so-called 
generalized Schwarzschild disks as their lowest order representatives.

The potential corresponding to a Kuzmin-Curzon disk is obtained by
introducing a discontinuity in the original Curzon metric 
(\cite{curzon24}; \cite{chazy24}) using the
 transformation $z \rightarrow |z|+b$. This way of introducing 
the discontinuity is equivalent to placing  two 
mirror particles of mass $M$  at a distance $b$ below and above
the $z=0$ plane on the $z$ axis. The Newtonian potential 
for the Kuzmin-Curzon disk is 
\begin{equation}
{\nu_o^{K-C}}=-{M \over \left[\rho^2+(|z| +b)^2\right]^{1\over 2}}
\label{potential}\,,
\end{equation}
and following Weyl (1917), one identifies it with the metric function
$\nu_o$ in equation (\ref{newmet}). This metric function 
remains a solution of the
Laplace equation outside the $z=0$ plane, but now due to
 the discontinuity, there appears to be a non-vanishing surface
 mass density. Integration of the remaining metric function yields 
\begin{equation}
{\zeta_o^{K-C}}=-{M^2 \rho^2 \over \left[\rho^2+(|z|+b)^2 \right]^2}\,.
\end{equation}
Correspondingly, as shown by BLK, the classical surface mass density
 $\Sigma(\rho)=2\,{\nu_o}_{,z}{|}_{z=0-}^{z=0+}$ is 
\begin{equation}
\Sigma^{K-C}={4\,M\,b \over  \left(\rho^2+
b^2\right)^{3\over 2}}\,.
\label{class}
\end{equation}
Here $M$ is the total mass of the disk as measured from the infinity,
and $b$ is a parameter measuring the compactness of the disk. 

The analogue of a generalized Schwarzschild disk is constructed
in the classical Kuzmin's picture by substituting the two mirror
particles by two rods with constant line density. Then, the general 
relativistic solution is given by 
\begin{equation}
{\nu_o^{G-S}}={\frac{M}{b_{max}-b_{min}}}
\log \left| \frac{\rho_{min}+| z |+b_{min}}{\rho_{max}+
| z |+b_{max}}\right|\,,
\end{equation}
and
\begin{equation}
{\zeta_o^{G-S}}=\frac{2\,M^2}{(b_{max}-b_{min})^2}
\log \left| \frac{(\rho_{min}+\rho_{max})^2-(b_{max}-
b_{min})^2}{4\,\rho_{min}\,\rho_{max}}\right|\,.
\label{potential3}
\end{equation}
In this case, the classical surface mass density $\Sigma(\rho)$ is 
\begin{equation}
\Sigma^{G-S}=\left.\frac{4\,M}{\left(b_{max}-
b_{min}\right)\left(\rho^2+
b^2\right)^{1\over 2} }\right|_{b_{max}}^{b^{min}},
\label{class2}
\end{equation}
where $\rho_{min}^2=\rho^2+(|z|+b_{min})^2$ and
 $\rho_{min}^2=\rho^2+(|z|+b_{min})^2$,
 $b_{max}-b_{min}$ being the parameter that measures the compactness
 of the disk.

\subsection{Disks in an expanding FRW Universe}
We start by looking somewhat more carefully at the stress-energy 
tensor for the solutions given by equation (\ref{generic}) 
with the metric functions corresponding to a generic thin
 counterrotating disk.
 
In order to deal with the discontinuities in the metric
across the $z=0$ plane, we consider the metric functions 
in the sense of distributions and introduce a new variable
 $\xi=|z|$.  Generically, the metric coefficients will have
 square-integrable weak derivatives, and so the usual formula
 for the Ricci tensor can be interpreted in the sense of 
distributions. The stress-energy tensor reads
 (cf. as well Chamorro, Gregory \& Stewart, 1987):
\begin{eqnarray}
 T_{\rho}^{\rho}-T_{z}^{z}&=&2\,R(t)^{-2}\left(
{\rho}^{-1}{\zeta}_{,\rho}+{\nu}_{,\xi}^{2}-{\nu}_{,\rho}^{2} 
\right)e^{-2\,\zeta+2\,{\nu}} \label{tensor1}\,,\\
T_{\rho}^{\rho}+T_{z}^{z}&=&-2\,R(t)^{-2}\left( {\dot R}^2
 +2 R {\ddot R}
 \right)e^{-2\,{\nu}}\label{tensor2}\,,\\
T_{t}^{t}+T_{\phi}^{\phi}&=&-2\,R(t)^{-2}\left(
{\nu}_{,\rho \rho}+{\nu}_{,\xi \xi}+2\,\delta(z)\,{\nu}_{,\xi}+
{\rho}^{-1}{\nu}_{,\rho}
-{\zeta}_{,\rho \rho}
-{\zeta}_{,\xi \xi}\right. 
\nonumber \\ &-&\left.2\,\delta(z)\,{\zeta}_{,\xi}-
{\nu}_{,\rho}^2- {\nu}_{,\xi}^2+
 \right)e^{-2\,\zeta+2\,U}\nonumber \\
&-&2\,\,R(t)^{-2}\left(2\,{\dot R}^2+
R {\ddot R} \right)e^{-2\,{\nu}}\label{tensor3}\,,\\
T_{t}^{t}-T_{\phi}^{\phi}&=&-2\,R(t)^{-2}\left(
{\nu}_{,\rho \rho}+{\nu}_{,\xi \xi}+2\,\delta(z)\,{\nu}_{,\xi}+
{\rho}^{-1}{\nu}_{,\rho}
 \right)e^{-2\,\zeta+2\,{\nu}}\nonumber \\
&-&2\,\,R(t)^{-2}\left({\dot R}^2-
R {\ddot R} \right)e^{-2\,{\nu}}\label{tensor4}\,,\\
T_{z}^{\rho}&=&-2\,R(t)^{-2}\left[\theta(z)-\theta(-z)\right]\left(
2\,{\nu}_{,\rho}{\nu}_{,\xi}-{\rho}^{-1}{\zeta}_{,\xi}
 \right)e^{-2\,\zeta+2\,{\nu}}\label{tensor5}\,,\\
T_{t}^{\rho}&=&2\,R(t)^{-3}{\dot R}\,{\nu}_{,\rho}
\,e^{-2\,\zeta+2\,{\nu}}\label{tensor6}\,,\\
T_{t}^{z}&=&2\,R(t)^{-3}\left[\theta(z)-\theta(-z)\right]{\dot R}
\,{\nu}_{,\xi}
\,e^{-2\,\zeta+2\,{\nu}}.
\label{tensor7}
\end{eqnarray}
The energy-momentum tensor given by equations 
(\ref{tensor1}-\ref{tensor7}) splits into a regular 
 and a singular part proportional to the functional $\delta(z)$,
 which is interpreted as a thin counterrotating disk.  
 We still separate the regular part into two bits: a highly
 inhomogeneous part vanishing at infinity that we interpret as an
interaction term between the matter composing the disk
and  the cosmic fluid, and another part representing at infinity
 an isotropic perfect fluid. 
Phenomenologically we write
\begin{equation}
 T_a^b= T_a^{b\,disk}+T_a^{b\,int}+e^{-2\,{\nu}}T_a^{b\,FRW}
\label{three}\,,
 \end{equation} 
where 
\begin{equation}
T_t^{t\,FRW}=3\,\,R(t)^{-2}\,{\dot R}^2\,,
\end{equation}
and 
\begin{equation}
T_{\phi}^{\phi\,FRW}=T_{\rho}^{\rho\,FRW}=
T_{z}^{z\,FRW}=-2\,\,R(t)^{-2}\left({\dot R}^2+
2\,R {\ddot R} \right).
\end{equation}
The identification of the other terms is straightforward.
The term we have referred to as ${T}_{a}^{b\,disk}$ is precisely
the one  which is discontinuous across the $z=0$ plane and which
 therefore gives rise to the non vanishing mass density. 
One has to be careful when interpreting equation (\ref{three}), for it
has a purely phenomenological character. Obviously none of the three
 terms in the r.h.s. of equation (\ref{three}) can be regarded as a
true energy-momentum tensor, since they do not satisfy the energy
 conservation equation $T_{\; ;b}^{a b}=0$ separately. Note, that
this decomposition into three terms is similar to that proposed in
 \S 2.2 based on  the character of the gradient of the scalar field.
Both approaches qualitatively  lead to the same physical description.

%-----------------------------------------------------------------%
%-----------------------------------------------------------------%
%-------------------Galactic dynamics-----------------------------%
%-----------------------------------------------------------------%
%-----------------------------------------------------------------%
 
\section{Dynamics of expanding disks} 
In this Section we study the kinematical quantities of relativistic
 disks and discuss the influence of the expansion. We explicitly
prove that the mass-energy density decreases with time, and also address
 the question of the effect of the 
expansion on the flattening of the rotation curves. Special attention
 will be paid to the comparison between 
a static disk and its counterpart in a pressure-free Universe. 
Our motivation is that the Universe, as observed at present, has
 negligible pressure. 

We now look at the surface mass density and streaming velocities of 
the particles on the disk. The $\tau_a^b$ surface components are 
obtained by integration across the disk of the $T_a^b$ components of 
the energy-momentum tensor,
\begin{equation}
\tau_a^b=\int T_a^b\, R(t)\,e^{(\zeta-\nu)}dz\,.
\end{equation}

As shown by BLK, the surface rest mass density in a fixed reference
 system reads $\sigma_0=2\sigma_p (1-v^2)^{-1/2}$, while the surface
 mass density in the same reference system is
 $\sigma=2\sigma_p (1-v^2)^{-1}=\sigma_0 (1-v^2)^{-1/2}$,
where $v$ is the rotational velocity and $\sigma_p$ is
 the proper rest mass density of one stream.

We first calculate the surface mass density
 ${\sigma}(\rho,t)$  on the disk plane by integrating the $T_t^t$
 term across the disk. Using equations 
(\ref{tensor3},\ref{tensor4}) one gets
\begin{equation} 
{\sigma}(\rho,t)=-\tau_t^t=-\int_{0-}^{0+}
 {{T_t}^{t}R(t)e^{(\zeta-\nu)}dz}= R(t)^{-1}e^{(\nu-\zeta)}(4
\,\nu_{,\xi}-2\,{\zeta}_{,\xi}){|}_{\xi=0}\,.
\end{equation} 
Substituting ${\zeta}_{,\xi}$ as calculated from
 equation (\ref{ein4}) we obtain after some algebra
\begin{equation}
{\sigma}(\rho,t)=
4\,R(t)^{-1}\,{\nu}_{,{\xi}}\,e^{(\nu-\zeta)}(1-
\rho\,f(\gamma)\,\nu_{,\rho}){|}_{\xi=0}\,,
\label{energy}
\end{equation} where we denote
$f(\gamma)=(3\,\gamma+3)/(3\,\gamma+5)$ and
$\gamma$ is the adiabatic index of the perfect fluid. 
Note that the static case will be recovered in the limit
 $\gamma \rightarrow \infty$, $f(\gamma)\rightarrow 1$. The
 particular case of a dust filled Universe
corresponds to $\gamma=0$ and $f(\gamma)=3/5$.

By integrating the $T_{\phi}^{\phi}$ term obtained
 from equations (\ref{tensor3},\ref{tensor4}) one gets
 \begin{equation}
 v^2{\sigma}={\tau}_{\phi}^{\phi }=
\int_{0-}^{0+}{ {T_\phi}^{\phi }R(t)e^{(\zeta -\nu
)}dz}=2\,R(t)^{-1}{e^{(\nu-\zeta)}}f(\gamma)\,
{\zeta,}_{\xi}{|}_{\xi=0} \,,
\end{equation}
or alternatively
\begin{equation}
v^2{\sigma}={4\,R(t)^{-1}{e^{(\nu-\zeta)}}\rho\,f(\gamma)\,
{\nu,}_{\rho}{\nu,}_{\xi}}|_{\xi=0} 
\label{speed} \,.
\end{equation}
 Then, from equations (\ref{energy},\ref{speed}) one readily 
obtains the expression for the square of the streaming velocities
\begin{equation}
 v^2=\left.{\rho\,f(\gamma)\,\nu_{,\rho}\over{1- \rho\,f(\gamma)
\,\nu_{,\rho}}} \right|_{\xi=0}\,.\label{velocity}
\end{equation} 

In order to emphasize the difference in the flattening of the rotation
 curves in the static and expanding cases we also evaluate the 
derivative of the velocity in the radial direction obtaining
\begin{equation}
v_{,\rho}=-\left.\frac{f(\gamma)^{1/2}\nu_{,\,\xi \xi}}
{2\,[{\nu,}_{\rho}
(1-\rho\,f(\gamma)\,{\nu,}_{\rho})]^{1/2}}\right|_{\xi=0}.
\end{equation}
 
Since the classical mass density $\Sigma(r)=4 
\,{\nu}_{,\xi}|_{\xi=0}$ of the disks  is positive everywhere, 
in order to ensure the positivity of the surface energy density 
$\sigma$ we must impose $\rho\,f(\gamma)\,\nu_{,\rho}{|}_{\xi=0}<1$,
as deduced from equation (\ref{energy}). This is 
equivalent to the fulfillment of the weak energy condition (cf. BLK).
 For the particular case of an expanding Kuzmin-Curzon disk this
 restriction reduces to 
\begin{equation}
\frac{M\,f(\gamma)}{b}< \frac{\sqrt 27}{2}\,,
\end{equation}
 whereas for
a generalized Schwarzschild disk it becomes
 \begin{equation}
\frac{M\,f(\gamma)}{(b_{max}-b_{min})}<
\frac{[1-(b_{min}/b_{max})^2]^{\frac{1}{2}}}
{[1-(b_{min}/b_{max})^{\frac{2}{3}}]^{\frac{3}{2}}}\,,
\end{equation}
therefore, the expanding disks may be denser than their static
counterparts.

Moreover,  if $\rho\,f(\gamma)\,\nu_{,\rho}{|}_{\xi=0}<1/2$ the
 dominant energy-condition holds and  the rotation velocities are
 subluminal. It is clear that as long as the dominant
 energy condition holds the positivity of the mass density is also
 ensured. 

To compare the streaming velocities of a static disk and any of its
time-evolving counterparts, one may look at the expression of
 their quotient, which reads
\begin{equation}
\frac{v^2(\gamma)}{v^2(\gamma=\infty)}=\left.f(\gamma){1-\rho\,
\nu_{,\rho}
\over{1- \rho\,f(\gamma)\nu_{,\rho}}}\right |_{\xi=0}.
\end{equation} 
Since $f(\gamma)\le 1$, we conclude that 
for expanding disks the velocities  are typically everywhere  lower 
than those for a static one (see Figure 1).

To compare our results with those obtained by BLK we represent
the kinematical quantities as functions of the {\it 
circumferential} radius $\rho_c=\rho\, e^{-\nu}$, where
$2\pi \rho_c$ is the physical circumference of a circle with
 radius $\rho=constant$. 
Note that for the solutions studied in this paper
 the comoving radius does not depend on the adiabatic index 
$\gamma$.

Yet, more interesting conclusions are obtained from the
 quotient between the velocity derivatives in the radial direction,
\begin{equation}
\left.\frac{v_{,\rho}(\gamma)}{v_{,\rho}(\gamma=\infty)}=
f^{1/2}(\gamma)\left({1-\rho\,\nu_{,\rho}
\over{1- \rho\,f(\gamma)\nu_{,\rho}}}\right)^{3/2}\right |_{\xi=0}.
\label{derive}
\end{equation}
From equation (\ref{derive}) it can be learnt that the slope of the
rotation curves for expanding disks is less pronounced than
 for static ones. It can be concluded that the expansion accentuates
 the flattening of the rotation curves (see Figure 2).

Another question to dwell on is the effect of the expansion 
on the mass-energy density on the disk plane. 
Similarly, as done above with the velocities and their radial 
derivatives, we look at the
 quotient of the mass energy density of an evolving and a static disk:
\begin{equation}
\frac{\sigma(\gamma)}{\sigma(\gamma=\infty)}=\frac{1-\rho\,f(\gamma)
\nu_{,\rho}}{1-\rho\,\nu_{,\rho}}\,{(t\,e^{\zeta_0})}^{-\frac{2}
{3\gamma+3}}\,.
\end{equation}
Since $\zeta_0$ is negative 
everywhere on the disk, we conclude that 
 expanding disks are denser than static ones,
with additional mass concentrated outside the center, 
whereas the central density is not affected by the expansion. 
Time-evolution of this quantity crucially depends on the adiabatic
 index $\gamma$. To be more specific, the mass-density dilution rate 
increases, keeps constant or decreases with time, if the expansion
 parameter $\gamma$ is smaller, equal or larger than $-1/3$ 
respectively.

 It is also interesting  to examine the surface rest mass 
density of the rotating particles as measured in a frame attached 
to fixed axes. This quantity is given by 
\begin{equation}
{\sigma}_0=\sigma\,\left(1-v^2\right)^{1/2}.
\end{equation}
Substituting the expression for $\sigma$ and $v$ in equations 
(\ref{energy},\ref{velocity}) we get
\begin{equation} 
{\sigma}_0(\rho,t)=4\,\nu_{,\xi}\,{e^{(\nu-\zeta)}}
[1-\rho\,f(\gamma)\,\nu_{,\rho}]^{\frac{1}{2}}
[1-2\,\rho\,f(\gamma)\,\nu_{,\rho}]^{\frac{1}{2}}
|_{\xi=0}\,.
\end{equation} 
The latter quantity is everywhere positive provided the dominant 
energy condition holds. Taking the previous results into account
 it is straightforward to see that the influence of the expansion
 on the surface rest mass density $\sigma_0$ 
 is the same as for the mass density $\sigma$.  

We have further defined a new function
$\Delta \sigma/\sigma\equiv \big[\sigma(\gamma)-
\sigma(\gamma=\infty)\big]/\sigma(\gamma)$ which 
gives the contrast in surface mass density 
between an expanding and a static disk. 
Figures 3a and 3b represent the contrast function 
$\Delta \sigma/\sigma$ and Figures 4a and 4b give their rest 
counterparts $\Delta \sigma_0/\sigma_0$ for both a generalized 
Schwarzschild disk and a Kuzmin-Curzon disk. It can be learnt from the
 figures that the expanding disks with the same mass parameter $M$
 are much denser than their static counterparts.

Finally, it is worth looking at the specific angular momentum of
 the counterrotating components of the disk. Bearing in mind that 
the specific angular momentum of a particle with 
rest mass $m$ rotating at radius $r$ is defined as
\begin{equation}
j=(p_{\phi \,\phi} / m)=g_{\phi\,\phi}\,d\,\phi /d \lambda,
\end{equation}
where $\lambda$ is proper time, then
\begin{equation}
j(\rho,t)=\left.R(t){\,\rho\,v\,e^{-\nu}\over{(1-
v^2)^{\frac{1}{2}}}}\right|_{\xi=0},
\end{equation}or alternatively
\begin{equation}
j(\rho,t)=\left.R(t){(\rho^3\,f(\gamma)\,
\nu_{,\rho})^{\frac{1}{2}}\,e^{-\nu}
\over{(1-2\,\rho\,f(\gamma)\,\nu_{,\rho})^{\frac{1}{2}}}}
\right|_{\xi=0}.\end{equation}
Similarly to what  happened with the rotation velocities, a smaller
 $\gamma$ gives a smaller angular momentum, which increases with time.
This behaviour is presented in Figures 5a and 5b where we compare
the specific angular momentum in an expanding and a static 
case for a generalized Schwarzschild disk and a Kuzmin-Curzon
disk.

\section{Conclusions and outlook}
We have constructed exact solutions to Einstein equations which we
interpret as representing relativistic disks in a cosmological 
setting. A self interacting scalar field serves as a source of the
 global expansion, yet locally it defines a disk-like 
structure across the $z=0$ plane. Far away from this plane the 
 gradient of the scalar field may be identified with the velocity
potential of an irrotational perfect fluid with an adiabatic equation
of state $p=\gamma \rho$. Near the azimuthal plane the spatial 
gradients of the scalar field dominate over the kinetic part and the
 geometry is highly inhomogeneous. No sharp-cut transition region
 exists in between these two different regimes, however, this probably 
corresponds to a physically more realistic situation than a surface
 matching between different solutions of gravitational field 
equations. Moreover, to the best of our knowledge, there do not 
exist such solutions in the case of axial symmetry and an
external FRW geometry.

Some particular cases of the solutions we have obtained ($k^2=6$)
may be re-interpreted as disks in the Brans-Dicke theory. Indeed, 
in this case the scalar field may be considered as massless (the
 potential term vanishes) and a simple conformal transformation 
$ds^2\rightarrow e^{\psi/{\sqrt {\omega +3/2}}}ds^2$
transforms the solution into a Brans-Dicke frame (\cite{taub73}).
The physical interpretation of the solutions then is quite 
different: these represent relativistic disks in a theory  where
the gravitational constant varies in space and time. We will not 
dwell more about this point here, but just mention that the analysis 
of our paper applies equally well to Brans-Dicke disks.

Once the solutions are interpreted as local inhomogeneities in a 
model Universe the way is cleared to see the effects produced by
expansion. We have found that depending on the rate of expansion
 the inhomogeneities occupy larger or smaller regions
 of the Universe. More specifically, for the accelerated expansion 
we find that inhomogeneities disappear with time, while in a
deccelerated model their growth is unbounded.

The effect of the expansion on the kinematical characteristics of the 
disks was studied as well. We have shown that expansion changes in
principle the fall-off of the rotational curves, the angular momentum
and the surface mass density of the disks. We have also compared the
characteristics of the static disks with those in a dust filled 
Universe.

Although we have concentrated our study on disk-like objects, it is
remarkable that the generating technique used to obtain solutions is
also adequate to study other type of sources of astrophysical interest
such as cosmic strings, walls, spherical shells, etc.

\section{Acknowledgements}
We are grateful to Wyn Evans, Sasha Kashlinsky, Konrad Kuijken
 and M. A. V\'azquez-Mozo for correspondence and valuable suggestions. 
This work was partially supported  by a Spanish Ministry of 
Education Grant (CICyT) PB93-0507 and a Basque Country University
 Grant UPV/EHU/72.310EBO36/95. R.L. acknowledges financial support 
from the Basque Goverment under Fellowship BFI94-094.

\clearpage
\begin{figure}[b]
{\noindent
\begin{minipage}[t]{.46\linewidth}
\begin{center}
\epsscale{0.95}
\plotone{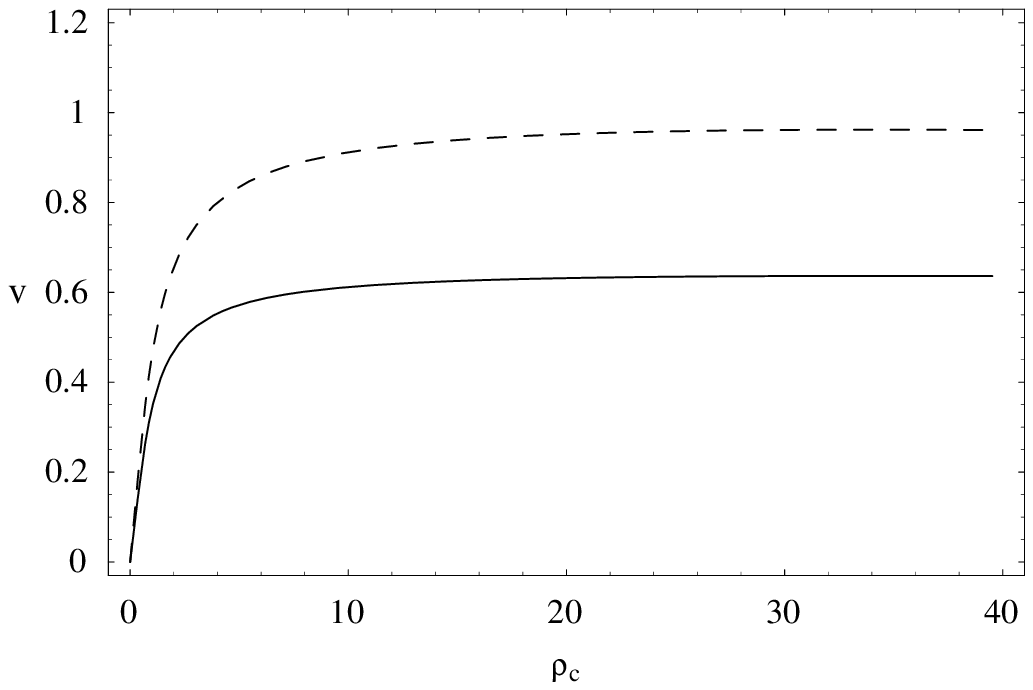}
\end{center}
\end{minipage}\hfill
\begin{minipage}[t]{.46\linewidth}
\begin{center}
\epsscale{0.97}
\plotone{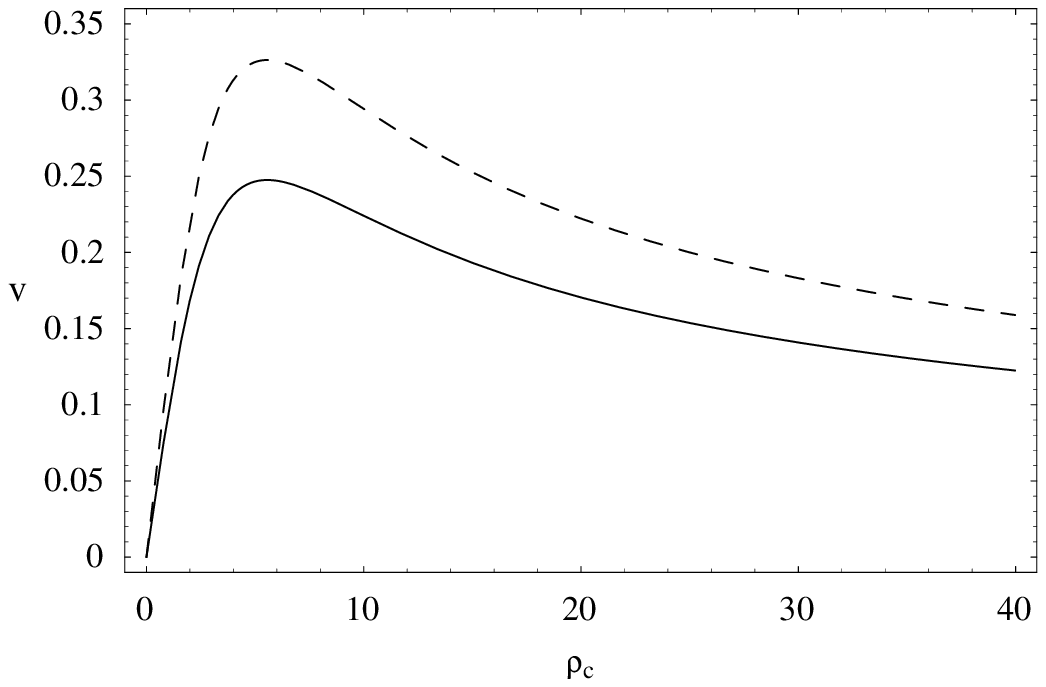}
\end{center}
\end{minipage}}
{\noindent
\begin{minipage}[t]{.46\linewidth}
\begin{center}
Fig. 1a
\end{center}
\end{minipage}\hfill
\begin{minipage}[t]{.46\linewidth}
\begin{center}
Fig. 1b
\end{center}
\end{minipage}}
\caption{Streaming velocity $v$ as 
function of the circumferential radius $\rho_c$  for a 
generalized Schwarzschild disk with $M=100$, $b_{min}=1$ 
and $b_{max}=200$ (a); and for a Kuzmin-Curzon disk with $M=1$
and $b=4$ (b). The continuous lines correspond to expanding disks
in a dust filled FRW Universe,whereas the dashed lines correspond
to their static counterparts.}
\end{figure}

\clearpage

\begin{figure}[b]
{\noindent
\begin{minipage}[t]{.46\linewidth}
\begin{center}
\epsscale{0.95}
\plotone{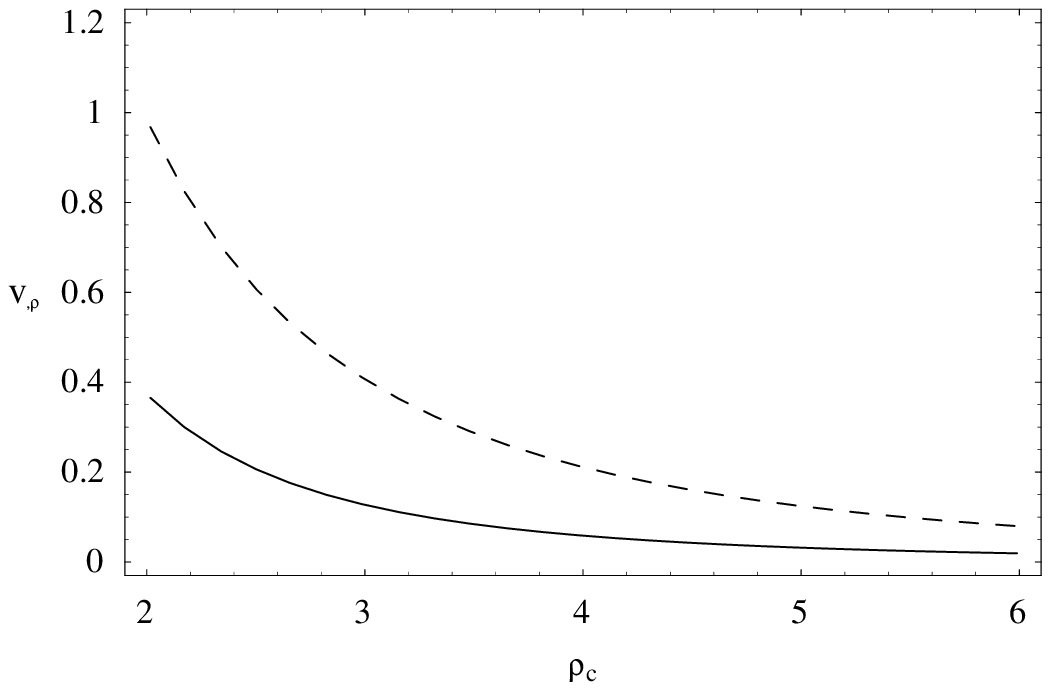}
\end{center}
\end{minipage}\hfil
\begin{minipage}[t]{.46\linewidth}
\begin{center}
\epsscale{0.95}
\plotone{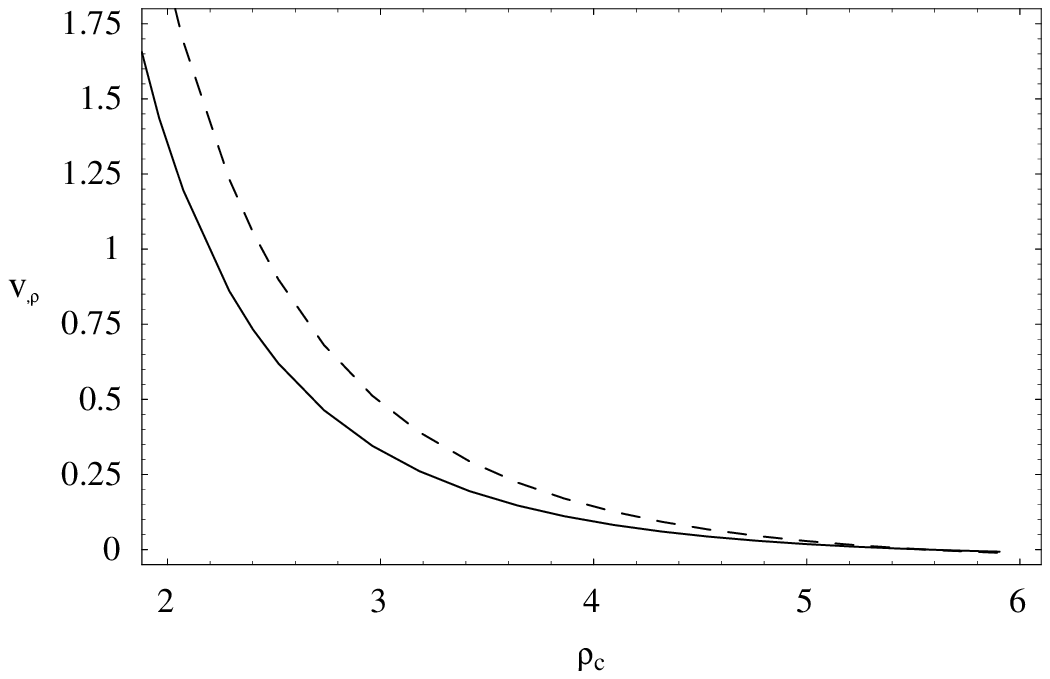}
\end{center}
\end{minipage}}
{\noindent
\begin{minipage}[t]{.46\linewidth}
\begin{center}
Fig. 2a
\end{center}
\end{minipage}\hfill
\begin{minipage}[t]{.46\linewidth}
\begin{center}
Fig. 2b
\end{center}
\end{minipage}}
\caption{ Radial derivative of the streaming velocity 
$v_{\rho}$ as function of the circumferential radius $\rho_c$  for 
a generalized Schwarzschild disk with $M=100$, $b_{min}=1$ 
and $b_{max}=200$ (a); and a Kuzmin-Curzon disk with $M=1$ and $b=4$
 (b).The continuous lines corresponds to expanding disks in a 
dust filled FRW Universe, whereas the dashed lines correspond to their
 static counterparts.}
\end{figure}
\clearpage
\begin{figure}[b]
{\noindent
\begin{minipage}[t]{.46\linewidth}
\begin{center}
\epsscale{0.95}
\plotone{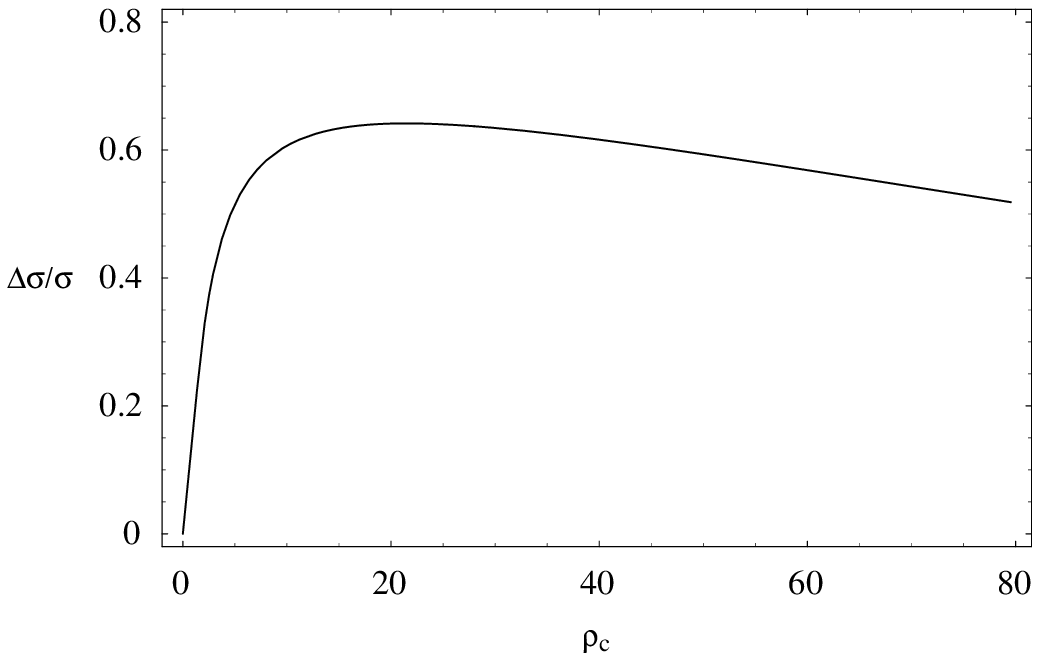}
\end{center}
\end{minipage}\hfill
\begin{minipage}[t]{.46\linewidth}
\begin{center}
\epsscale{0.95}
\plotone{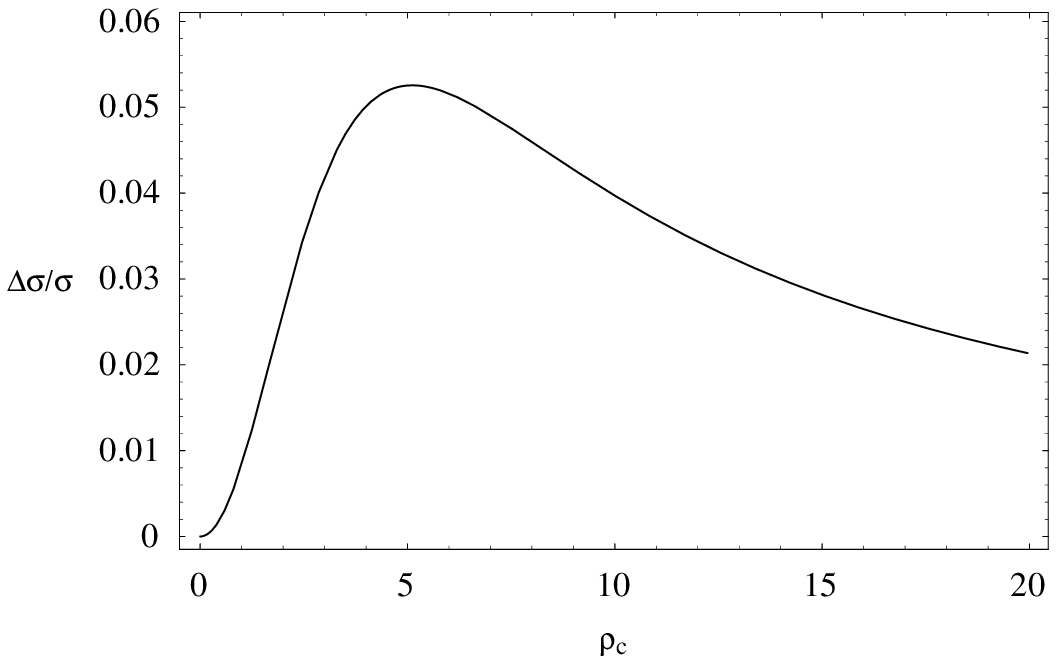}
\end{center}
\end{minipage}}
{\noindent
\begin{minipage}[t]{.46\linewidth}
\begin{center}
Fig. 3a
\end{center}
\end{minipage}\hfill
\begin{minipage}[t]{.46\linewidth}
\begin{center}
Fig. 3b
\end{center}
\end{minipage}}
\caption{Surface energy density contrast $\Delta \sigma/\sigma$ 
at $t=1$ as function of the circumferential radius $\rho_c$ for a
 generalized Schwarzschild disk with $M=100$, $b_{min}=1$ 
and $b_{max}=200$ (a); and for Kuzmin-Curzon disk with 
$M=1$ and $b=4$ (b).}
\end{figure}
\clearpage
\begin{figure}[b]
{\noindent
\begin{minipage}[t]{.46\linewidth}
\begin{center}
\epsscale{0.95}
\plotone{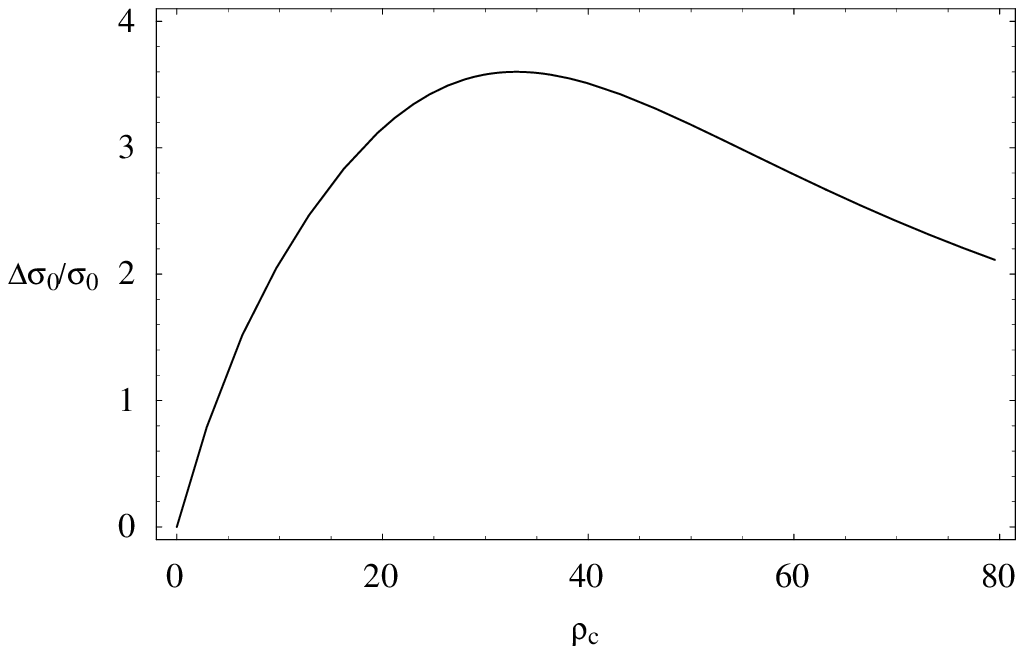}
\end{center}
\end{minipage}\hfill
\begin{minipage}[t]{.46\linewidth}
\begin{center}
\epsscale{0.95}
\plotone{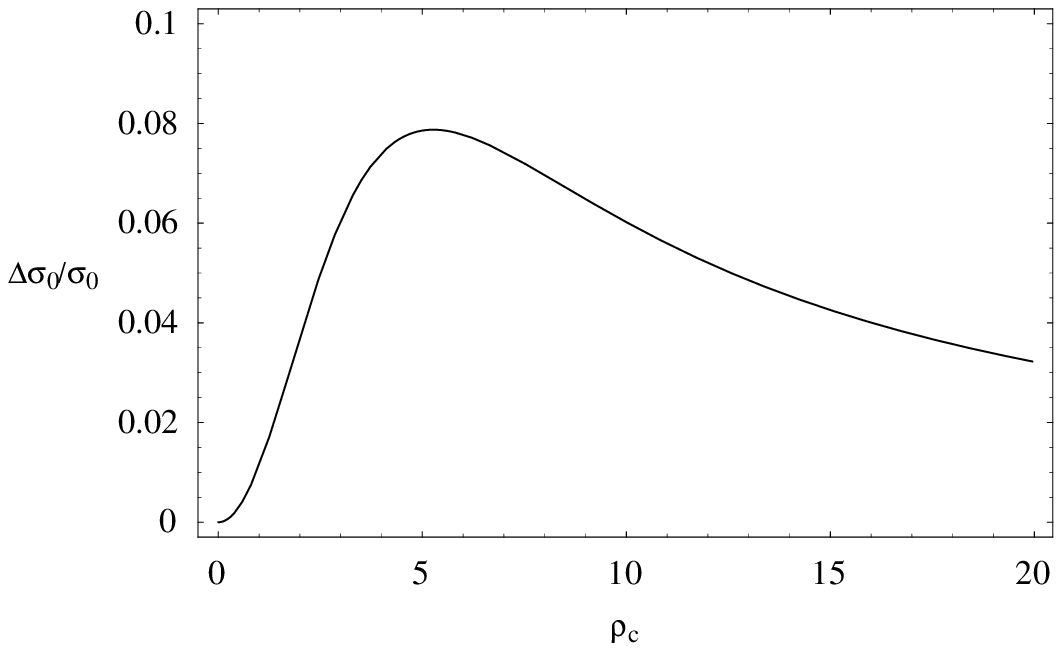}
\end{center}
\end{minipage}}
{\noindent
\begin{minipage}[t]{.46\linewidth}
\begin{center}
Fig. 4a
\end{center}
\end{minipage}\hfill
\begin{minipage}[t]{.46\linewidth}
\begin{center}
Fig. 4b
\end{center}
\end{minipage}}
\caption{Surface rest energy density contrast $\Delta\,
 \sigma_0/\sigma_0$ at $t=1$ as function of the circumferential radius 
${\rho}_c$  for a generalized Schwarzschild disk with $M=100$,
 $b_{min}=1$ and $b_{max}=200$ (a); and a Kuzmin-Curzon disk with
 $M=1$ and $b=4$ (b).}
\end{figure}
\clearpage
\begin{figure}[b]
{\noindent
\begin{minipage}[t]{.46\linewidth}
\begin{center}
\epsscale{0.95}
\plotone{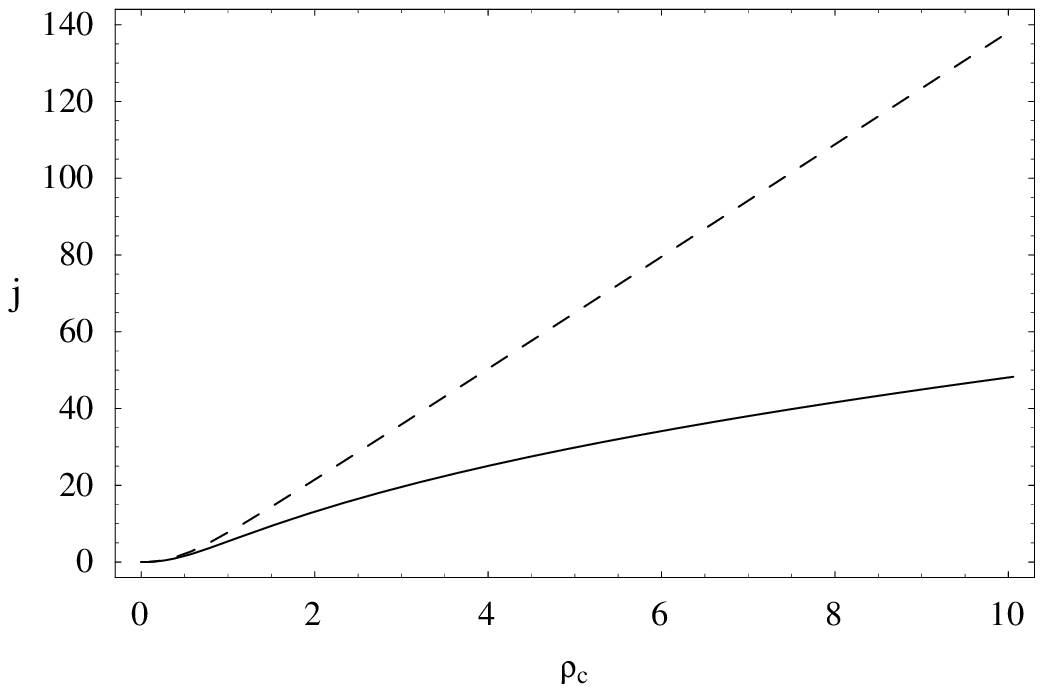}
\end{center}
\end{minipage}\hfill
\begin{minipage}[t]{.46\linewidth}
\begin{center}
\epsscale{0.95}
\plotone{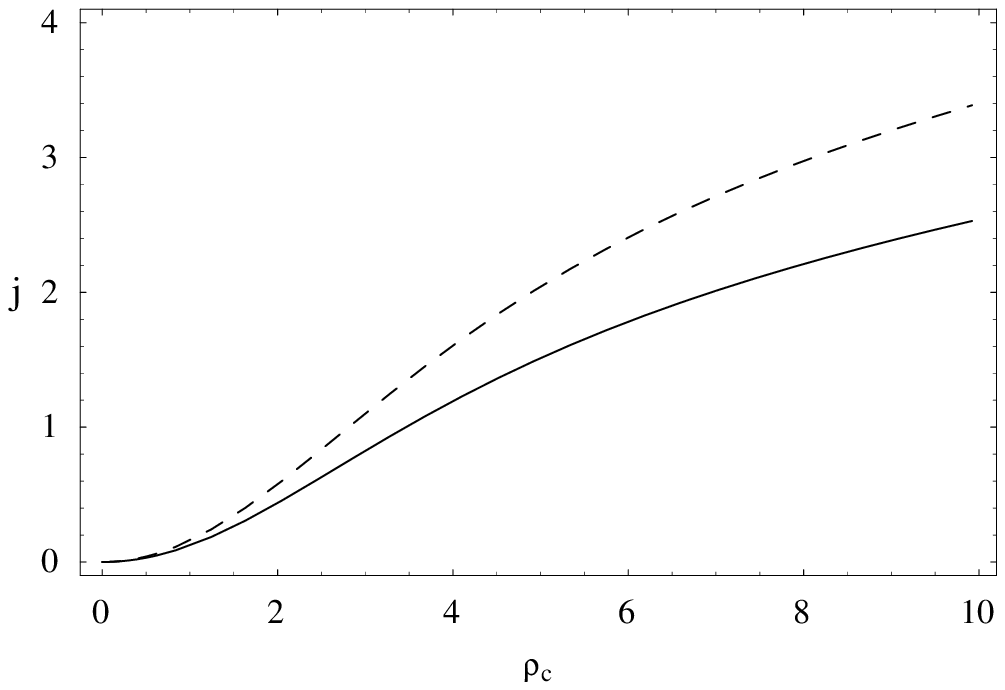}
\end{center}
\end{minipage}}
{\noindent
\begin{minipage}[t]{.46\linewidth}
\begin{center}
Fig. 5a
\end{center}
\end{minipage}\hfill
\begin{minipage}[t]{.46\linewidth}
\begin{center}
Fig. 5b
\end{center}
\end{minipage}}
\caption{Specific angular momentum $j$ at $t=1$ as 
function of the circumferential radius $\rho_c$  
for a generalized Schwarzschild disk with $M=100$, $b_{min}=1$ 
and $b_{max}=200$ (a); and for a Kuzmin-Curzon disk with $M=1$ 
and $b=4$ (b). The continuous lines correspond to disks in a 
dust filled FRW Universe, whereas the dashed lines correspond
 to their static counterparts. }
\end{figure}


\begin{thebibliography}{}
\bibitem[Bertola et al. 1996]{bertola96} Bertola, F., Cinzano, P., 
Corsini, E.M., Pizzella, A., Persic, M., \& Salucci, P. 1996,
 \apj, 458, L67

\bibitem[1993]{ledvinka93} Bi\u{c}\'ak, J., Ledvinka, T. 1993, 
Phys. Rev. Lett. 71, 1669

\bibitem[1993]{bijak93} Bi\u{c}\'ak, J., Lynden-Bell, D., 
\& Katz, J.  1993, Phys. Rev. D 47, 4334  


\bibitem[1993]{pichon93} Bi\u{c}\'ak, J., Lynden-Bell, D.,
 \& and Pichon, C.  1993, \mnras, 265, 126

\bibitem[Curzon 1924]{curzon24} Curzon, H. E. J. 1924, Proc. London
 Math. Soc. 23, 477

\bibitem[Chamorro, Gregory, \& Stewart 1987]{cha87} Chamorro, A.,
 Gregory, R., \& Stwewart, J. M. 1987, Proc. R. Soc. London A, 413, 251

\bibitem[Chazy 1924]{chazy24} Chazy, J. 1924, Bull. Soc. Math. 
Paris 52, 17 

\bibitem[1992]{evans92} Evans, N. W., \&  de Zeeuw, P. T. 1992, 
\mnras, 257, 152

\bibitem[1995]{fein95} Feinstein, A., Ib\'a\~nez, J.,
 \& Lazkoz, R. 1995, Class. Quantum. Grav., 12, L57
 
\bibitem[1995]{fon95} Fonarev, O. A. 1995, Class. 
Quantum. Grav., 12, 1739

\bibitem[Kalnajs 1976]{kalnajs76} Kalnajs, A., 1976, \apj, 205, 751 

\bibitem[Kuijken, Fisher \& Merrifield 1996]{konrad96} Kuijken, K., 
Fisher D., \& Merrifield, M. R.. 1996, \mnras, 283, 543

\bibitem[Kuzmin 1956]{kuzmin56} Kuzmin, G. G. 1956, Astron. Zh., 
33, 27 

\bibitem[Lemos \& Ventura 1994]{lemos94} Lemos, J. P. S., 
\& Ventura, O. S. 1994, J. Math. Phys., 35, 3604

\bibitem[1993]{letelier93} Lemos J. P. S., \& Letelier, P. S. 1993,
 Class. Quantum Grav., 10, L75

\bibitem[1994]{letelier94} 
 Lemos J. P. S., \& Letelier, P. S. 1994, Phys. Rev. D, 49, 5135

\bibitem[1996]{letelier96} 
 Lemos J. P. S., \& Letelier, P. S. 1996,
 Int. J. Mod. Phys. D, 5, 53 

\bibitem[Levi-Civita 1919a]{levi19a} Levi-Civita, T. 1919a, 
Rend. Acad. Lincei, 28, 3

\bibitem[Levi-Civita 1919b]{levi19b} Levi-Civita, T. 1919b, 
Rend. Acad. Lincei, 28, 101

\bibitem[Noerdlinger \& Petrosian 1971]{petros71} Noerslinger, P.D.,
\& Petrosian, V. 1971, 168, 1

\bibitem[1996]{pichon96} Lynden-Bell, D., \& and Pichon, C.  1996,
 \mnras, 280, 1007

\bibitem[Merrifield \& Kuijken 1994]{konrad94} Merrifield, M.R., 
\& Kuijken, K. 1994, \apj, 432, 575

\bibitem[Mestel 1963]{mestel63} Mestel L. 1963, \mnras, 126, 553

\bibitem[1969a]{morgan69a} Morgan, T., \& Morgan, L. 1969a, 
Phys. Rev., 183, 1097

\bibitem[1969b]{morgan69b} Morgan, T., \& Morgan, L. 1969b, Phys. 
Rev., 188, 2544 (E)

\bibitem[1970]{morgan70} Morgan, T., \& Morgan, L. 1970, Phys. Rev.
 D, 2, 2576 

\bibitem[Nagai \& Miyamoto 1976]{nagai76} Nagai, R., \& Miyamoto,
 M. 1976, PASJ, 28, 1 

\bibitem[Rix, Franx, Fisher \& Illingworth 1992]{rix92} Rix, H.,
 Fisher, H., \& Illingworth, D. 1992, \apj, 400, L5

\bibitem[Tabensky and Taub 1973]{taub73} Tabensky, R., \& Taub, A.H. 
1973, Commun. Math. Phys. 29, 61  

\bibitem[Toomre 1963]{toomre63} Toomre, A., 1963, \apj, 138, 385 

\bibitem[Weyl 1917]{weyl17} Weyl, H. 1917, Ann. Phys. (N.Y.), 54, 307

\end{thebibliography}
\end{document}